\documentclass[twocolumn,article]{aastex631}
\usepackage{graphicx}
\usepackage{isotope}
\usepackage{amsmath}
\usepackage{orcidlink}

\begin{document}

\title{The Rotation Dip in the Envelope-Disk Transition of HH 111: Evidence for Magnetic Braking}

\author[0009-0006-0486-5468]{Jyun-Heng Lin}
\affiliation{Department of Physics, National Taiwan Normal University, Taipei 116, Taiwan; \href{mailto:jyunhenglin@asiaa.sinica.edu.tw}{jyunhenglin@asiaa.sinica.edu.tw}}
\affiliation{Academia Sinica Institute of Astronomy and Astrophysics, Taipei 106216, Taiwan, R.O.C.}
\affiliation{Taiwan Astronomical Research Alliance (TARA), Taiwan}

\author[0000-0002-3024-5864]{Chin-Fei Lee}
\affiliation{Academia Sinica Institute of Astronomy and Astrophysics, Taipei 106216, Taiwan, R.O.C.}
\affiliation{Taiwan Astronomical Research Alliance (TARA), Taiwan}

\author[0000-0002-7402-6487]{Zhi-Yun Li}
\affiliation{Astronomy Department, University of Virginia, Charlottesville, VA, USA}
\author[0000-0003-3497-2329]{Yueh-Ning Lee}
\affiliation{Department of Earth Sciences, National Taiwan Normal University, Taipei 116, Taiwan}
\affiliation{Center of Astronomy and Gravitation, National Taiwan Normal University, Taipei 116, Taiwan}
\affiliation{Physics Division, National Center for Theoretical Sciences, Taipei 106, Taiwan}

\author[0000-0002-4392-1446]{Ya-Lin Wu}
\affiliation{Department of Physics, National Taiwan Normal University, Taipei 116, Taiwan; \href{mailto:jyunhenglin@asiaa.sinica.edu.tw}{jyunhenglin@asiaa.sinica.edu.tw}}
\affiliation{Center of Astronomy and Gravitation, National Taiwan Normal University, Taipei 116, Taiwan}
\author[0000-0002-5845-8722]{J. A. López-Vázquez}
\affiliation{Academia Sinica Institute of Astronomy and Astrophysics, Taipei 106216, Taiwan, R.O.C.}




\begin{abstract}
Magnetic braking can drive angular momentum loss in star formation and influence disk evolution. A previous study of HH 111 VLA1 suggested a decrease in rotation velocity in a region between the infalling envelope and rotating disk. Using ALMA C$^{18}$O ($J = 2-1$) data, we analyzed the gas motion within 6000 au and found clear deviations from the simplest expectations of free-fall towards the central star with conserved angular momentum in the transition region between the envelope and disk (from $\sim$5200 to 160 au). The region can be further divided into three zones: (1) outer region with a significant decrease in infall velocity, dropping to approximately 60\% of the free-fall velocity and 70\% of conservation of angular momentum and energy; (2) middle region with a sharp drop in angular momentum and thus rotation velocity and an increase in infall velocity; and (3) inner region with rotation velocity increasing inward to connect to that of the Keplerian disk and infall velocity decreasing to zero.  Comparison with non-ideal MHD simulations suggests that the reduced infall velocity in the outer region can be due to magnetic tension by the pinched magnetic field lines, the sharp drop of angular momentum in the middle region can be due to magnetic braking as the field lines pile up, and the rapid increase in rotation velocity in the inner region might result from weaker magnetic braking due to ambipolar diffusion of the field lines. The resulting dip in the rotation profile supports magnetic braking.

\end{abstract}



\section{Introduction} \label{sec:intro}

Molecular clouds are magnetized \citep[e.g.,][]{Troland_2008ApJ...680..457T, Planck_2015A&A...576A.104P}, and recent polarimetric observations from the James Clerk Maxwell Telescope (JCMT) BISTRO survey \citep[e.g.,][]{Ward-Thompson_2017ApJ...842...66W,Pattle_2019ApJ...880...27P} have revealed the presence of organized magnetic fields in dense cores. These observational results indicate the need to include magnetic fields in realistic models, as they can significantly impact the dynamics of core collapse and star formation. In particular, magnetic fields can efficiently extract angular momentum from collapsing gas through magnetic braking \citep{Allen(b)_2003ApJ...599..363A,Mellon_2008ApJ...681.1356M}, thereby regulating the formation and extent of Keplerian disks \citep{Wurster_2018MNRAS.480.4434W}.

Numerical and theoretical studies \citep[e.g.,][]{Krasnopolsky_2011ApJ...733...54K, Li_2011ApJ, Wurster_2018FrASS...5...39W, Zhao_2020SSRv..216...43Z, Lee_YN_2021A&A...648A.101L} suggest that magnetic braking becomes especially efficient in the envelope-disk transition region (hereafter the transition region), where the infalling envelope gradually transforms into a rotationally supported disk. In these models, magnetic braking is found to be most effective when the mass-to-flux ratio is relatively low, typically with dimensionless values $\lambda \lesssim 5$ \citep[e.g.,][]{Mellon_2008ApJ...681.1356M, Li_2011ApJ}. In this region, magnetic field lines are pinched and twisted into toroidal components, generating magnetic torques that extract angular momentum and slow rotation. As a result, both the rotation and infall velocity profiles are modulated. According to simulations, a distinct dip is predicted in the rotation curve between the envelope and the Keplerian disk. The transition is also where the gas kinematics starts to deviate significantly from the simplest expectations of free-fall with conserved angular momentum, due to angular momentum loss through magnetic braking. 

Despite extensive theoretical work, direct observational constraints on magnetic braking and the transition region remain limited. During the past decade, numerous studies have investigated the kinematics of early protostellar systems. Observations with the Submillimeter Array \citep[SMA; e.g.,][]{Lee_2010ApJ} and the Atacama Large Millimeter/Submillimeter Array \citep[ALMA; e.g.,][]{Murillo_2013ApJ...764L..15M, Aso_2015ASPC..499..285A, Aso_2017ApJ...849...56A} have revealed a transition from $v_\phi\sim r^{-1}$ (corresponding to the outer envelope) directly to $v_\phi\sim r^{-0.5}$ (corresponding to the Keplerian disk).

\defcitealias{Lee_2016ApJ}{Paper I}

\citet{Lee_2010ApJ} first introduced the possibility of a transition region in HH 111 VLA 1, which was later confirmed by ALMA observations in \citet{Lee_2016ApJ} (hereafter Paper I). \citetalias{Lee_2016ApJ} reported that the specific angular momentum decreases by a factor of $\sim$3 in the region between $\sim$2000 and 160 au, reaching a minimum at a radius of about 800 au. However, detailed kinematics has not been well quantified. In this study, we follow up the work of \citetalias{Lee_2016ApJ} by resolving the velocity profile in the transition region to strengthen the observational evidence for magnetic braking. We construct a kinematic model to characterize angular momentum loss and the associated changes in rotation and infall velocities, and compare our results with theoretical predictions. 

The HH 111 system, located in the L1617 cloud of Orion at a distance of $\sim$400 pc \citep{Zucker_2019ApJ...879..125Z}, consists of two sources: VLA 1 and VLA 2. Our target, VLA 1, is a young Class I protostar with a flattened envelope. The total mass of the protostar and disk is estimated to be $\sim$1.8$M_\odot$, consisting of $\sim$1.22$M_\odot$ for the protostar and $\sim$0.58$M_\odot$ for the disk \citep{Lee_2024ApJ...971L..23L}. The envelope mass is estimated to be $\sim$0.6$M_\odot$ from $\sim$160 au (0$\overset{\prime\prime}{.}$4) to 8000 au (20$^{\prime\prime}$), and the systemic velocity in the LSR frame is 8.85 km s$^{-1}$, both reported in \citetalias{Lee_2016ApJ}. Based on the inside-out collapse model \citep{Shu_1977ApJ...214..488S}, the infall radius is estimated to be $\sim$30000 au. This estimate adopts a sound speed of $\sim$0.27 km s$^{-1}$, corresponding to a temperature of 20 K that is typical of observed Class I protostellar envelopes \citep[e.g.,][]{Jorgensen_2002A&A...389..908J,Launhardt_2013A&A...551A..98L}, and an age of $5 \times 10^5$ yr \citep{Lee_2020NatAs...4..142L}. These values suggest that the initial collapsing core has extended well beyond the currently observed envelope. The disk is nearly edge-on, with an inclination angle of 15$^\circ$–18$^\circ$ relative to the line of sight \citep{Lee_2020NatAs...4..142L}, which is almost perpendicular to the inner jet axis \citep{Reipurth_1992ApJ}, and the position angle (P.A.) is $\sim$6$\overset{\circ}{.}$7.

Our observations are presented in Section \ref{sec:observation}. The results, discussion, and conclusion are presented in Section \ref{sec:result}, \ref{sec:Disscussion} and \ref{sec:conclusion}, respectively.

\section{Observations} \label{sec:observation}
The HH 111 system was observed using ALMA Band 6 with the 12m array, 7m array, and total power dishes. Most data were obtained during Cycle 1 (Project ID: 2012.1.00013.S) from 2013 to 2015, with additional 12m data from Cycle 4 (Project ID: 2016.1.00389.S) in 2016. We used the C$^{18}$O ($J = 2$–$1$) line from the 230 GHz receiver to trace the envelope and disk structures (see Table 1 for details).

To simultaneously present both extended and compact structures, the 12m and 7m data were combined using the tclean task in the Common Astronomy Software Applications (CASA) package \citep{CASA_2022PASP..134k4501C}. For Cycle 1, the calibrators included J0750+1231 and J0607–0834 as bandpass calibrators, J0532+0732 and J0607–0834 as gain calibrators, and Callisto and Ganymede as flux calibrators. For Cycle 4, J0750+1231 and J5010+1800 were used as bandpass calibrators, J0552+0313 as the gain calibrator, and J0750+1231 and J0423–0120 as flux calibrators. Robust weighting parameters of 0.5 and -2 for the visibilities yielded synthesized beam sizes of 0$\overset{\prime\prime}{.}$73 × 0$\overset{\prime\prime}{.}$61, and 0$\overset{\prime\prime}{.}$16 × 0$\overset{\prime\prime}{.}$13, respectively. To recover the missing flux, the feather task \citep{Plunkett_2023PASP..135c4501P} was used to incorporate total power measurements into the final map. The resulting C$^{18}$O channel maps have rms noise levels of approximately 4.47 mJy beam$^{-1}$ (robust parameter = 0.5) and 2.62 mJy beam$^{-1}$ (robust parameter = -2), with a velocity resolution of 0.15 km s$^{-1}$.

\begin{deluxetable*}{lccccccc}[!t]
\tabletypesize{\scriptsize}
\tablewidth{10pt}
\tablenum{1}
\tablecaption{Observation information}
\tablehead{
\colhead{Project ID} & \colhead{Array} & \colhead{Execution Block} & \colhead{Observation date} & \colhead{Total integral time} & \colhead{Configuration} &\colhead{Baseline range} 
} 
\startdata 
2012.1.00013.S & 12m & 3 & Apr 13-28, 2014 & 140 min& C32-4&20-558.2 m\\ 
& 7m & 20 & Dec 14, 2013-Dec 14, 2014 & 445 min & 7-m & 7-48.9 m\\
& TP & 8 & Aug 26 - Sep 4, 2015 & 266 min & TP & None\\
\hline
2016.1.00389.S & 12m & 4 & Oct 6 - Oct 8, 2016 & 159 min& C40-6 & 17-3247 m
\enddata
\end{deluxetable*}

\section{Results}\label{sec:result}
The HH 111 protostellar system consists of an extended infalling envelope, a Keplerian disk, and a transition region. The envelope, which extends to $\sim$20$^{\prime\prime}$ ($\sim$8000 au), has a mass of $\sim$0.6$M_\odot$ \citepalias{Lee_2016ApJ}, about one third of the combined mass of the central protostar and disk \citep[$\sim$1.8$M_\odot$;][]{Lee_2024ApJ...971L..23L}. In the outer envelope, the gas is nearly quasi-static, as discussed earlier in Section \ref{sec:intro}. Within the Keplerian disk ($<$160 au or $0\overset{\prime\prime}{.}4$), the gas motion exhibits Keplerian rotation. Between the infalling envelope and the rotationally supported disk, a transition region is expected to form under the influence of a non-negligible magnetic field in the core, as the material gradually loses angular momentum and settles into Keplerian rotation. This angular momentum loss is caused by magnetic field lines transporting angular momentum away from the gas in this region, as shown in previous simulations \citep{Li_2011ApJ,Zhao_2016MNRAS.460.2050Z}.

\begin{figure*}[t]
    \includegraphics[width=\linewidth]{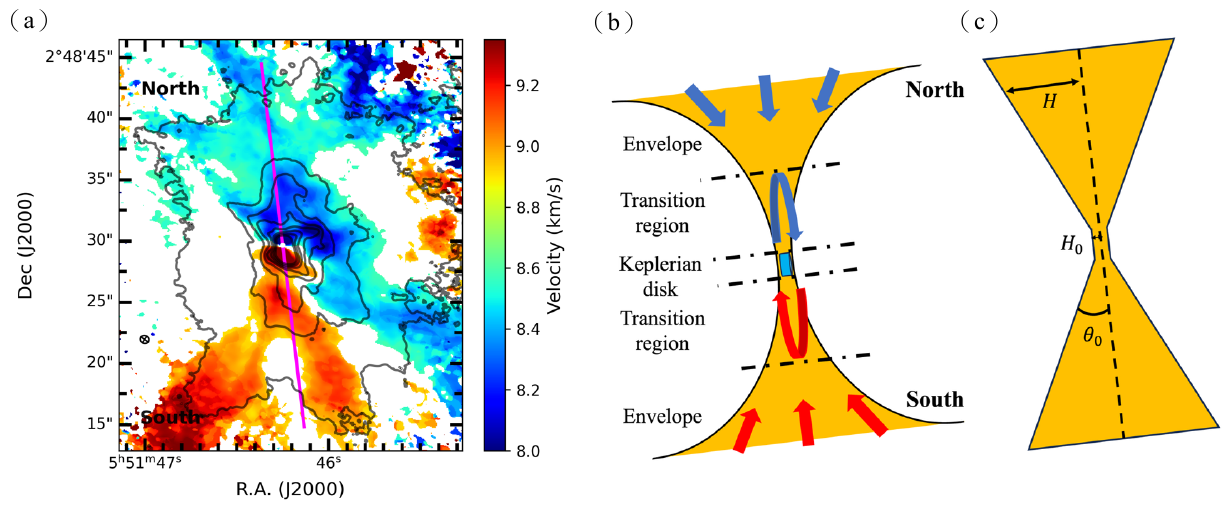}
    \caption{Left panel shows the flux-weighted velocity map (moment 1) of C$^{18}$O ($J=2-1$) in HH 111, integrated over the velocity range from 7 to 12 km s$^{-1}$. The contours represent the total intensity map (moment 0), starting at 2$\sigma$ and increasing in steps of 3$\sigma$ ($\sigma$ = 28 mJy beam$^{-1}$ km s$^{-1}$) up to 25$\sigma$. The magenta line indicates the PV cut, taken along the disk major axis at a P.A. of $\sim$6.7$^\circ$, as determined from the observed disk orientation. The middle panel illustrates the structural distribution of an edge-on protostellar system, and the right panel shows a schematic of the simplified flattened envelope model; a detailed description is given in Section \ref{sec:model}.}
    \label{fig:moment}
\end{figure*}
\subsection{A Flattened Structure of the Envelope}

Left panel of Figure~\ref{fig:moment} shows the total intensity map (moment 0) on the flux-weighted velocity map (moment 1) of HH 111, derived from C$^{18}$O ($J = 2$–$1$) emission integrated from 7 to 12 km s$^{-1}$. The intensity map (contours) reveals a flattened envelope with a size of $\sim$30$^{\prime\prime}$, while the velocity map (color) displays a clear velocity gradient: blueshifted emission in the northern region and redshifted emission in the southern region. The schematic diagram in the middle of Figure~\ref{fig:moment} follows \citetalias{Lee_2016ApJ}, illustrating an edge-on system composed of an envelope ($\gtrsim$2000 au), a transition region ($\sim$2000 to 160 au), and a Keplerian disk ($\lesssim$160 au), with the northern side blueshifted, the southern side redshifted, and a P.A. of $\sim$6$\overset{\circ}{.}$7, along which both the velocity gradient and the density structure are aligned.

\subsection{Rotation and Infall Motion}\label{sec:rot_inf}

\begin{figure}[h]
\centering
    \includegraphics[width=\linewidth]{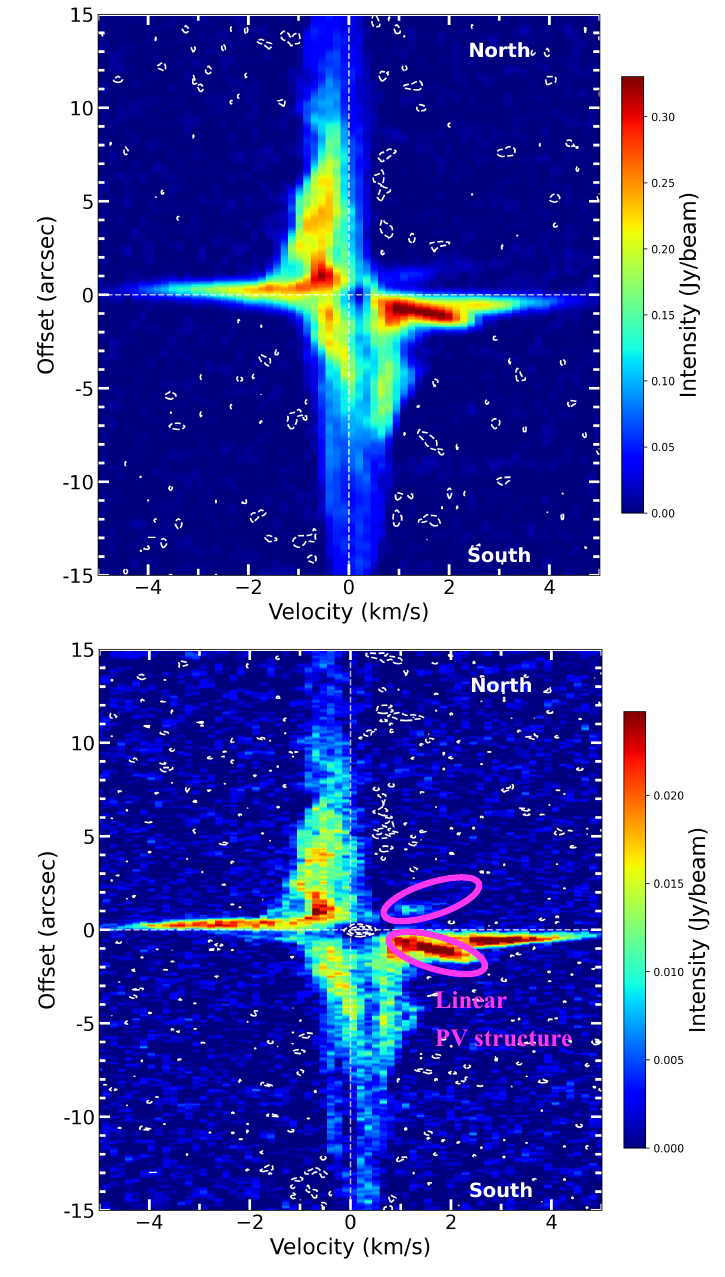}
    \caption{The C$^{18}$O PV diagrams are cut along the major axis of the disk. The x-axis represents the velocity relative to the $V_{sys}$ = 8.85 km s$^{-1}$, while the y-axis shows the positional offset from the source center. Top panel shows the PV diagram obtained from the channel maps with a robust parameter of 0.5, while bottom panel from the channel maps with a robust parameter of -2. The corresponding angular resolutions are 0$\overset{\prime\prime}{.}$73 × 0$\overset{\prime\prime}{.}$61 and 0$\overset{\prime\prime}{.}$16 × 0$\overset{\prime\prime}{.}$13, respectively.}
    \label{fig:pv_obs}
\end{figure}

Figure~\ref{fig:pv_obs} presents the position-velocity (PV) diagram of the C$^{18}$O ($J = 2-1$) line, obtained by making a cut along the major axis of the disk (P.A. = 6$\overset{\circ}{.}$7). The cut extends 15$^{\prime\prime}$ ($\sim$6000 au) from the protostar to the north and south into the envelope (as indicated by the magenta line in Figure~\ref{fig:moment}). Two PV diagrams are shown using two different robust weightings of visibilities: robust parameter of 0.5 (top) and robust parameter of -2 (bottom) for low (0$\overset{\prime\prime}{.}$73 × 0$\overset{\prime\prime}{.}$61) and high (0$\overset{\prime\prime}{.}$16 × 0$\overset{\prime\prime}{.}$13) angular resolutions, respectively.

In the low-resolution PV diagram, the line of sight velocity of the extended envelope increases gradually from $\sim$15$^{\prime\prime}$ ($\sim$6000 au) to 5$^{\prime\prime}$ ($\sim$2000 au) along both the northern and southern extensions. A sharp velocity drop is then seen from $\sim$3$^{\prime\prime}$ (1200 au) to 1$^{\prime\prime}$ (400 au), as pointed out in \citetalias{Lee_2016ApJ}. In contrast, the high-resolution PV diagram reveals a sharp increase in velocity from approximately 0.5 to 4 km s$^{-1}$ inside $\sim$0$\overset{\prime\prime}{.}$5, corresponding to the Keplerian disk.

Additionally, we detected negative flux in the PV diagram, as indicated by white dashed contours in Figure~\ref{fig:pv_obs}. In the south, a negative feature near 0.5 km s$^{-1}$ extends from $\sim$6$^{\prime\prime}$ to 15$^{\prime\prime}$ in the high-resolution data; a similar feature is found in the north between $\sim$0.6 to 0.7 km s$^{-1}$, up to $\sim$15$^{\prime\prime}$ north of the center. These features suggest the presence of absorption, which could be caused by gas falling inward. Moreover, absorption at this velocity is consistent with the infall velocity derived later in the model (see Section \ref{sec:Disscussion}), further supporting the interpretation.

We also observed two linear PV structures, highlighted by pink ellipses in Figure~\ref{fig:pv_obs}. One extends from the center of the system out to $\sim-2^{\prime\prime}$, at $\sim$1 to 2 km s$^{-1}$, consistent with \citetalias{Lee_2016ApJ}; a faint northern counterpart is also seen within a similar velocity range, extending from the center to $\sim$2$^{\prime\prime}$.
\subsection{A Flattened Infalling Rotating Envelope Model} \label{sec:model}
To interpret the C$^{18}$O ($J = 2-1$), we constructed a flattened, infalling, and rotating envelope model based on \citet{Lee_2010ApJ} and extended it to include the transition region and Keplerian disk. The model combines structural profiles (half-thickness, density, temperature) with kinematic prescriptions (rotation and infall velocities):

The half-thickness of the flattened envelope, $H$, is expressed in cylindrical coordinates ($R$, $\theta$, $z$) as:
\begin{equation}
H(R) = \text{max}\left( H_0, R\text{ tan}\theta_0\right),
\end{equation}
where $H_0$ is the minimum of half-thickness and $\theta_0$ is the flaring angle measuring from the mid-plane (See the right panel of Figure~\ref{fig:moment} for a schematic illustration of the model structure). The number density of molecular hydrogen in the envelope is given by
\begin{equation}
n(R,z)=n_0 \left(\frac{\sqrt{R^2+z^2}}{R_0}\right)^p,
\end{equation}
where $R_0$ is 2000 au (5$^{\prime\prime}$), $n_0$ is mid-plane number density at $R_0$, and $p$ is a power-law index. The C$^{18}$O abundance relative to molecular hydrogen is fixed at 1.7 $\times$ 10$^{-7}$. The temperature is \textbf{similarly} described by:
\begin{equation}
T(R,z)=T_0 \left(\frac{\sqrt{R^2+z^2}}{R_0}\right)^q,
\end{equation}
where $T_0$ is the temperature at $R_0$ and \textit{q} is the power-law index. We adopt single power-law forms across the envelope, transition region, and disk. Although simplified, this approach enables a focused analysis on velocity structure. The resulting discrepancy between the model emission intensity and the observed emission intensity will be discussed later. 

In the envelope, we assume a ballistic motion under the conservation of angular momentum and energy \citep{Sakai_2014Natur.507...78S}. In this model, the total energy of the infalling gas is assumed to be zero, under the assumption that collapse begins from large distances where both the gravitational potential and initial velocities are negligible. Although the envelope has finite thickness, the radial and rotation velocities are assumed to be the same as their mid-plane values at each radius:
\begin{equation}
\label{eq.4}
v_{\phi}(R) = v_{\phi,\ell}(R) \equiv \frac{\ell_0}{R},
\end{equation}

\begin{equation}
\label{eq.5}
v_{r}(R) = v_{r,\ell,e}(R) \equiv \sqrt{\frac{2G(M_*+M_d)}{R}-v_{\phi,\ell}^2},
\end{equation}
where $\ell_0$ is the initial specific angular momentum at large distance before the infall, $G$ is the gravitational constant and $M_*+M_d=1.8M_\odot$ is the total mass of the protostar and disk in VLA 1. The mass of the envelope is $\sim$0.6$M_\odot$, which contributes less than 15\% to the infall velocity and is therefore neglected.

In the transition region, deviations from conservation are expected due to magnetic effects \citep{Li_2011ApJ,Zhao_2020MNRAS.492.3375Z,Lee_YN_2021A&A...648A.101L}. Since no analytic solution exists for the transition region, we adopted a power-law form as the angular momentum distribution for the first try:

\begin{equation}
\label{eq.6}
    \ell(R) = \ell_0\left(\frac{R}{R_e}\right)^{b},
\end{equation}
where $R_e$ is 2000 au and $b$ is the power-law index. This assumption provides a baseline for comparing the model with the observational data in more detail. Similar to Eq.~\ref{eq.4} and \ref{eq.5}, the rotation velocity can be derived from a power-law distribution of angular momentum, and the corresponding infall velocity can be obtained as follows:
\begin{equation}
\label{eq.7}
v_{\phi}(R) =  \frac{\ell(R)}{R},
\end{equation}

\begin{equation}
\label{eq.8}
v_{r}(R) = \sqrt{\frac{2G(M_*+M_d)}{R}-v_{\phi}^2}.
\end{equation}

The envelope is assumed to transition to a Keplerian disk at the disk radius $R_d$ of $\sim$160 au \citep{Lee_2020NatAs...4..142L}. In the Keplerian disk, the infall velocity vanishes and the rotation velocity becomes Keplerian:
\begin{equation}
\label{eq.9}
v_{\phi,\text{Kep}}(R) = \sqrt{\frac{G(M_*+M_d)}{R}},
\end{equation}
where $v_{\phi,\text{Kep}}$ is the Keplerian motion.

The model spans from 10 au (0$\overset{\prime\prime}{.}$025) to 8000 au (20$^{\prime\prime}$), covering the entire system. Radiative transfer was performed using RADMC-3D \citep{Dullemond_2012ascl.soft02015D} under LTE to produce model channel maps and then the PV diagram of C$^{18}$O. The inclination angle of the flattened envelope to the plane of sky is $\sim$18$^\circ$, a mean value derived from the disk \citep{Lee_2020NatAs...4..142L} and jet \citep{Reipurth_1992ApJ}. The position angle of the disk is $\sim$6$\overset{\circ}{.}$7 (\citetalias{Lee_2016ApJ}). The power-law index \textit{b} is determined using the envelope angular momentum $\ell_0 \sim 1333$ au km s$^{-1}$ \citep{Lee_2010ApJ} and the value at the disk radius $R_d$, which is derived from Keplerian rotation as $\sim$530 au km s$^{-1}$. These values yield \textit{b} $\sim$ 0.38. The best-fit physical parameters are: $H_0 \sim$ 0$\overset{\prime\prime}{.}$4 (160 au), $\theta_0 \sim 18^\circ$, $n_0 \sim 1.3 \times 10^6$ cm$^{-3}$, $p \sim -0.13\pm0.01$, $T_0 \sim 20\pm1$ K and $q \sim -0.5\pm0.1$. These parameters are obtained by adjusting those previously obtained from fitting the spectrum, moment map, and PV diagrams of C$^{18}$O at lower resolution \citep{Lee_2010ApJ}.

\begin{figure*}[t]
   \includegraphics[width=0.96\linewidth]{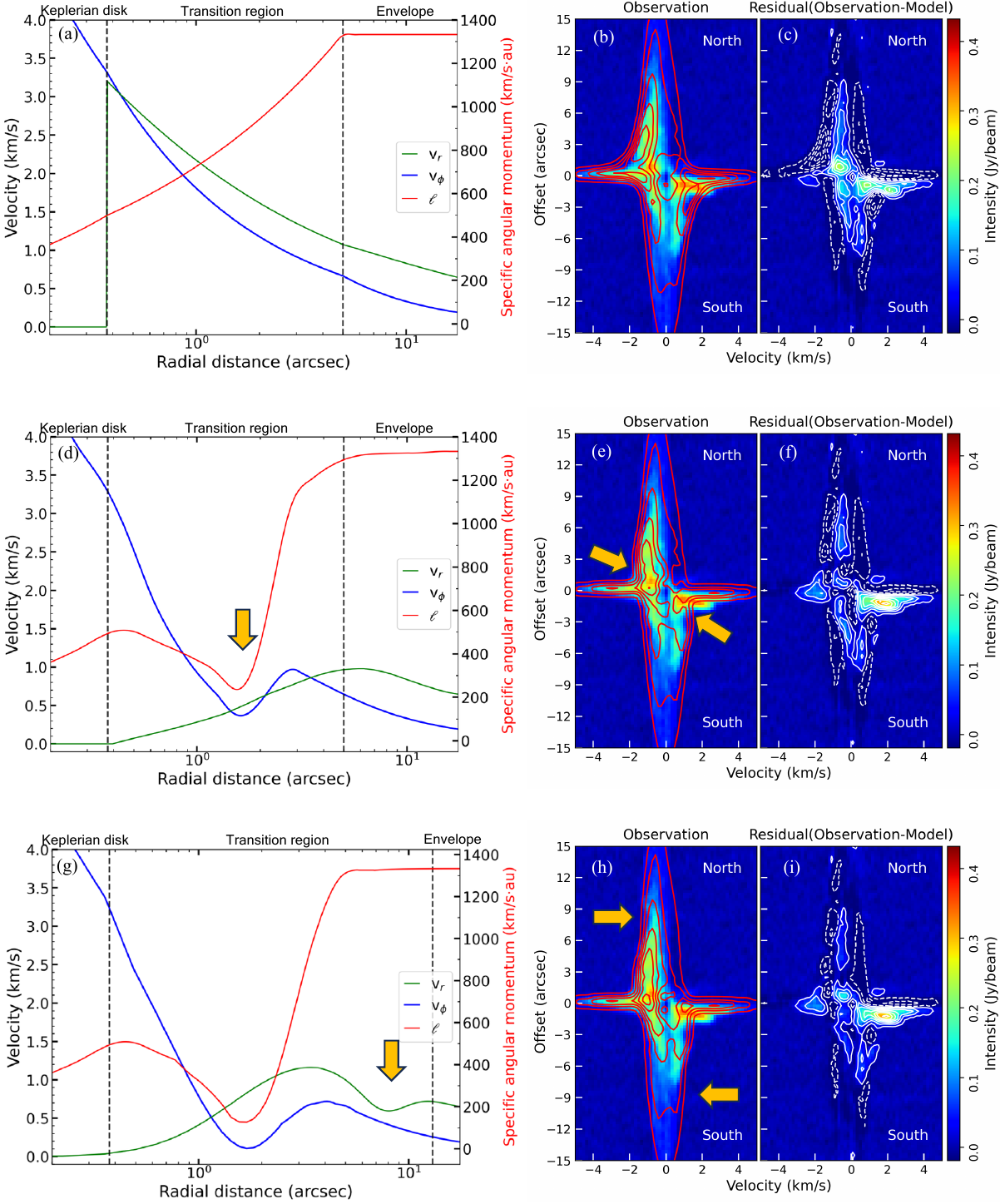}
    \caption{In each row, from left to right, the panels display the model velocity profiles, the model–observation comparison, and the residual map, respectively. In the velocity setup (a), (d), and (g), the red line represents the specific angular momentum $\ell$, with its values shown on the right-hand y-axis. The blue and green lines indicate the rotation velocity $v_\phi$ and infall velocity $v_r$, respectively. Two vertical black dashed lines divide the diagram into three parts: from the outer to the inner regions, they represent the envelope, the transition region, and the Keplerian disk. In the PV diagrams (b), (e), and (h), the color scale shows the observational data and the red contours the model, while in the residual maps (c), (f), and (i), the color scale and white contours show the data–model residuals. The red and white contours start at 4$\sigma$ and increase in steps of 10$\sigma$ ($\sigma$ = 4.8 $\times$ 10$^{-3}$ Jy beam$^{-1}$). An apparent asymmetry between the redshifted and blueshifted sides in the model is present, which is primarily due to self-absorption (see Section~\ref{sec:rot_inf}}).
    \label{fig:pv_obs_model}
\end{figure*}

Figure~\ref{fig:pv_obs_model}(a) presents the radial profiles of angular momentum and velocity. As discussed earlier, the velocity structure shown in the Figure is divided into three regions from the outside inward: the envelope, the transition region, and the Keplerian disk. In the envelope, we assume conservation of angular momentum and energy, and the rotation and infall velocities are derived using Eq.~\ref{eq.4} and \ref{eq.5}. In the transition region, the angular momentum is described by a power-law distribution (Eq.~\ref{eq.6}), and the corresponding rotation and infall velocities are computed using Eq.~\ref{eq.7} and \ref{eq.8}. In the inner Keplerian disk, the rotation follows Keplerian motion (Eq.~\ref{eq.9}), while the infall velocity approaches zero.

Figure~\ref{fig:pv_obs_model}(b) displays the model PV diagram as contour levels overlaid on the observed PV diagram, which was generated with a robust weighting of 0.5. The residual map in Figure~\ref{fig:pv_obs_model}(c), obtained by subtracting the model from the observed PV diagram, emphasizes the discrepancies between the two and illustrates the regions where the model deviates from the observations. While the outer envelope (outside $\sim$9$^{\prime\prime}$) and the disk (within $\sim$0$\overset{\prime\prime}{.}$5) are well reproduced, the intermediate region ($\sim$9$^{\prime\prime}$ to 1$^{\prime\prime}$) shows a sharp velocity drop of $\sim$70\% below the power-law prediction (Figure~\ref{fig:pv_obs_model}(c)). This discrepancy motivates the additional modeling of a rotation velocity dip, as discussed in the next subsection.
\subsection{A Dip in Rotation Velocity}

To address this sharp velocity decrease observed in the transition region, we introduced a modification to the model. A rotation dip is expected from theoretical and numerical studies \citep{Krasnopolsky_2011ApJ...733...54K, Li_2011ApJ, Zhao_2020MNRAS.492.3375Z, Lee_YN_2021A&A...648A.101L} because the pile-up of magnetic field lines in the transition region can lead to such a dip in rotation velocity due to non-ideal MHD effects. For example, the dotted line in the left panel of Figure 14 in \citet{Li_2011ApJ} illustrates this profile under the assumption of axisymmetry. Therefore, we adopted the angular-momentum–conserving profile as the baseline and introduced a log-Gaussian term to model the dip:

\begin{equation}
  v_{\phi}(R) = v_{\phi,\ell}(R) \,\bigg[ 1 - A_{\phi} \exp\!\left(-\frac{(\ln(R/R_{\phi,dip}))^2}{2\sigma_{\phi}^2}\right) \bigg],
\end{equation}
where $A_\phi$ is the depth of the dip (ranging from 0 to 1), $R_{\phi,dip}$ is the characteristic radius of the dip center, and $\sigma$ is the dimensionless width. It sets the radial extent of the dip relative to $R_{\phi,dip}$, with the half-maximum points given by $R/R_{\phi,dip} = \exp(\pm \sqrt{2\ln 2}\,\sigma)$. The width is different inside and outside the dip position and can be defined as:
\begin{equation}
  \sigma_{\phi} =
  \begin{cases}
      \sigma_{\rm in}, & R \leq R_{\phi,dip}\\[6pt]
      \sigma_{\rm out}, & R > R_{\phi,dip}
  \end{cases},
\end{equation}
The best-fit parameters are: $A_\phi$ $\sim$ 0.8, $R_{\phi,dip}$ $\sim$ 1$\overset{\prime\prime}{.}$5 (600 au), $\sigma_{\rm in}$ $\sim$ 1.6, and $\sigma_{\rm out}\sim$0.4. The logarithmic form better captures the scale-dependent nature of the dip than a linear Gaussian.

For the infall velocity, we refine it to have a simple gradual decline as our focus in this second attempt is to better reproduce the rotation velocity profile:
\begin{equation}
    \label{eq.12}
    v_r(R)= v_{r,\ell,e}(R_{c}) \exp\!\left(-\frac{(\ln(R/R_{c}))^2}{2\sigma_{c}^2}\right),
\end{equation}
where $R_c$ marks the starting point of the decline, and $\sigma_c$ is the width of the log-Gaussian profile. Here, $R_c\sim5^{\prime\prime}$(2000 au), $\sigma_c\sim1$. As shown in Figure~\ref{fig:pv_obs_model}(d), the infall velocity decreases smoothly from $\sim5^{\prime\prime}$ and approaches to zero near $R_d$.

This refinement significantly improved the agreement between the model and the observed velocity profiles, particularly in reproducing the dip between $\sim$4$^{\prime\prime}$ to 1$^{\prime\prime}$, which is indicated by a yellow arrow in Figure~\ref{fig:pv_obs_model}(e) and (f). However, discrepancies remain in the outer region ($\sim$13$^{\prime\prime}$ to 5$^{\prime\prime}$), where the observed PV diagram shows a $\sim$ 30\% drop (0.9 km s$^{-1}$ to 0.6 km s$^{-1}$) that is not well captured by the present velocity profiles. 

\subsection{A Dip in Infall Velocity}
The discrepancies in the outer region ($\sim$13$^{\prime\prime}$ to 5$^{\prime\prime}$) might be caused by the infall velocity because the rotation velocity is small there. Several numerical simulations \citep{Allen(a)_2003ApJ...599..351A, Li_2011ApJ, Zhao_2020MNRAS.492.3375Z} have predicted a decline in infall velocity due to magnetic tension (see also the dotted line in the left panel of Figure 14 in \citet{Li_2011ApJ}). Motivated by this, we introduced a similar log-Gaussian dip in the infall profile, applied only at radii greater than 3$^{\prime\prime}$:

\begin{equation}
    v_r(R)=v_{r,\ell,e}(R)\,\bigg[ 1 - A_{r} \exp\!\left(-\frac{(\ln(R/R_{r,dip}))^2}{2\sigma_{r}^2}\right) \bigg] ,
\end{equation}
where $A_r$, $R_{r,dip}$, $\sigma_r$ represent the depth of the dip, the characteristic radius of the dip, the width of the log-Gaussian profile, respectively.

Additionally, we shift the position of the peak infall velocity from $\sim$ 5$^{\prime\prime}$ to 3$\overset{\prime\prime}{.}$5 and change the profile to log-Gaussian as Eq.~\ref{eq.12}, in order to have a better match to the observations. Among various forms tested, a log-Gaussian again provided a simple yet effective match, applied only at radii smaller than 3$\overset{\prime\prime}{.}$5.

The best-fit parameters for infall velocity are: $A_r \sim$ 0.35, $R_{r,dip}\sim7\overset{\prime\prime}{.}5$ (3000 au), $\sigma_r\sim0.3$, $R_c\sim3\overset{\prime\prime}{.}5$ (1400 au), and $\sigma_c\sim0.85$. To further improve the fit, we also slightly refined the rotation velocity profile with $R_{\phi,dip}$ $\sim$ 1$\overset{\prime\prime}{.}$5 (600 au), $A_\phi$ $\sim$ 0.9, $\sigma_{\rm in}$ $\sim$ 1.3, and $\sigma_{\rm out}\sim$ 0.5.

Figure~\ref{fig:pv_obs_model}(g)–(i) show the updated model and residuals. The yellow arrows in panels (h) and (i) highlight the improved agreement in the outer envelope. As the infall velocity starts decreasing around 13$\arcsec$, we define this as the onset of the transition region, marked by the black dashed line in Figure~\ref{fig:pv_obs_model}(g)

\subsection{The Best-fit Model}
As shown in Figure~\ref{fig:pv_obs_model}(i), the residuals are reduced to below approximately 10\% of the original values in most regions. However, within 1$^{\prime\prime}$, they remain relatively large, with some areas retaining up to 50\%, likely due to the reported asymmetry in the innermost envelope \citep{Lee_2024ApJ...971L..23L} as well as the influence of the linear PV structures. Despite these localized discrepancies, the fitting (Figure~\ref{fig:pv_obs_model}(h) and (i)) matches observation well, with velocity uncertainties estimated to be less than 20\%.

Notice that some residuals should also result from the simplified power-law assumptions for density and temperature. However, the fitting results indicate that these assumptions do not affect the inferred velocity structure. We therefore adopt this as our best-fit model.

\begin{figure}[t]
\centering
    \includegraphics[width = 1\linewidth]{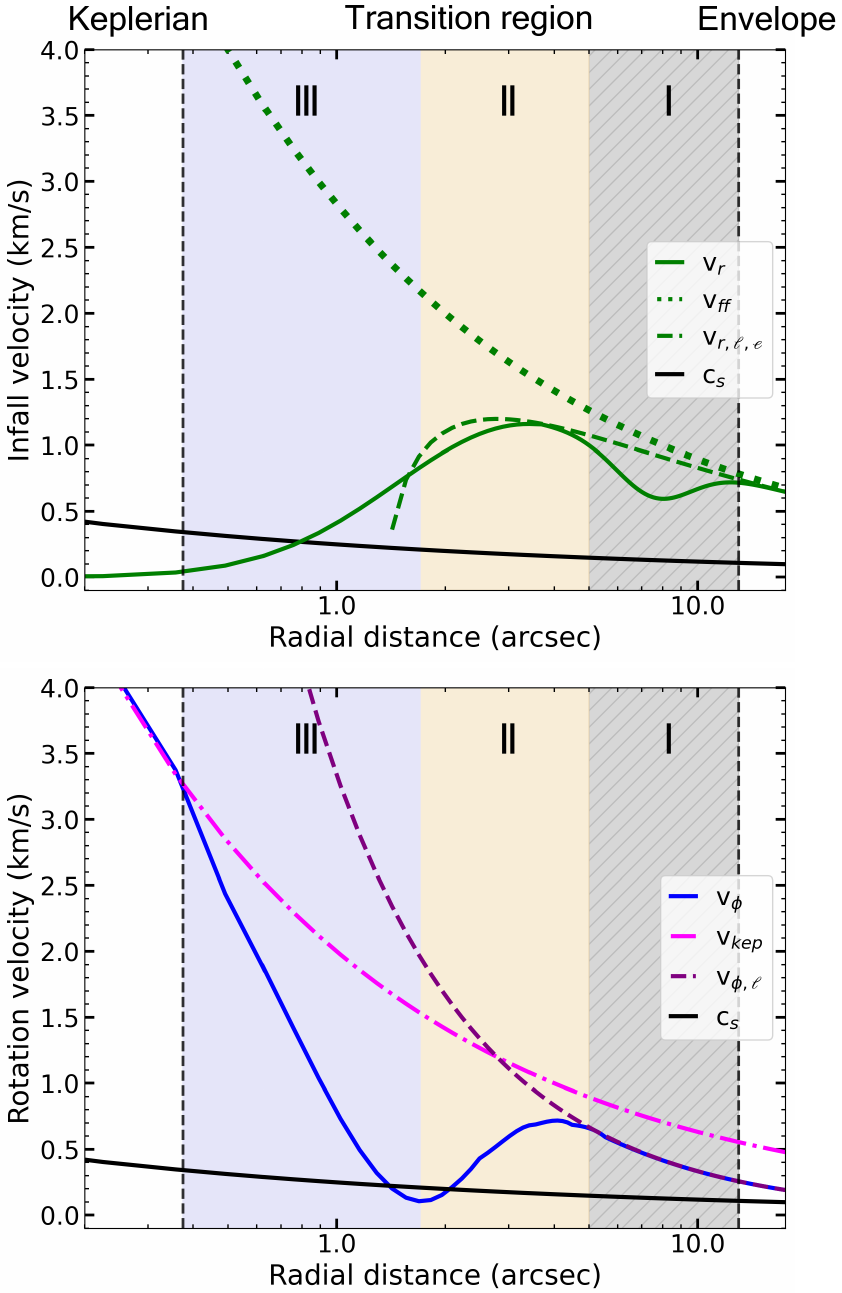}
    \caption{In the top panel, the infall velocity $v_r$ from the best-fit model (as in Figure~\ref{fig:pv_obs_model}(g)) is shown as a green solid line, while the green dotted line represents the free-fall velocity $v_{\text{ff}}$, and the green dashed line corresponds to the infall velocity derived from both angular momentum and energy conservation $v_{r, \ell,e}$. In the bottom panel, the rotation velocity $v_\phi$ used in the best-fit model is shown as a blue solid line. The pink dash-dotted line represents the rotation velocity derived from Keplerian motion $v_{\text{Kep}}$ the purple dashed line indicates the rotation velocity derived from specific angular momentum conservation $v_{\phi, \ell}$, and the black line represents the sound speed.}
    \label{fig:comparison}
\end{figure}

\section{Discussion} \label{sec:Disscussion}
\subsection{Comparison to theoretical Model}
Figure~\ref{fig:comparison} compares the rotation and infall velocity profiles of the best-fit model with free-fall, Keplerian rotation, sound speed, as well as the rotation velocity (Eq.\ref{eq.4}) and infall velocity (Eq.\ref{eq.5}) from the ballistic model. 
In the best-fit model, outside $\sim$13$^{\prime\prime}$ infall velocity follows the conserved-energy profile (green dashed line), remaining close to the free-fall velocity (green dotted line). However, inside this radius, it deviates sharply, dropping to $\sim$60\% of the free-fall value and $\sim$70\% of the conserved value near 7$^{\prime\prime}$. 
Between 7$^{\prime\prime}$ and 3$\overset{\prime\prime}{.}$5, it slightly recovers to $\sim$70\% of the free-fall value and remains close to the conserved value before steadily declining to zero at the Keplerian disk.

The rotation velocity agrees with angular momentum conservation for the region outside 5$^{\prime\prime}$. Around 4$^{\prime\prime}$, it begins to decrease significantly, reaching a minimum of 0.1 km s$^{-1}$ at 1$\overset{\prime\prime}{.}$5 with approximately 5\% of the initial conserved value. Inside this point, the rotation increases again and smoothly connects to the Keplerian profile at about 0$\overset{\prime\prime}{.}$4.

Based on these profiles, we identify three distinct kinematic zones in the transition region:

\subsubsection{Region I: Magnetic Cushion}
In Region I ($\sim$13$^{\prime\prime}$ to 5$^{\prime\prime}$, gray slashed shading), the infall velocity exhibits a clear dip, possibly due to magnetic tension, but its overall trend is still largely constrained by energy and angular momentum conservation. During the collapse, the magnetic field is initially dominated by the poloidal field. As the gas falls along the equatorial plane, the field lines are gradually dragged inward and become increasingly pinched, amplifying the outward magnetic tension that acts against infall \citep{Allen(a)_2003ApJ...599..351A}. 

\citet{Li_2011ApJ} adopted a moderate magnetic field strength corresponding to a dimensionless mass-to-flux ratio $\lambda\approx2.9$ and predicted a similar effect in simulations, where magnetic fields were found to be trapped in the post-shock region of an ambipolar diffusion-induced C-shock (see, e.g.,  the left panel of their Fig.~5), which reduced the infall velocity by at least 10\% compared to free-fall. In our case, the reduction is about $\sim$40\%, possibly resulting from a stronger initial magnetic field corresponding to a smaller dimensionless mass-to-flux ratio or differing physical conditions. The trend, however, remains qualitatively consistent.

Moreover, the symmetric absorption features observed in both the northern and southern extensions of the envelope provide further support for this interpretation. Their velocities are consistent with the infall profile in Region I, suggesting that the large-scale envelope is indeed undergoing infall, possibly moderated by magnetic pressure near the outer transition boundary.

\subsubsection{Region II: Magnetic Braking}
In Region II ($\sim$5$^{\prime\prime}$ to 1$\overset{\prime\prime}{.}$5, wheat shading), as the gas continues to collapse, the increasing rotation velocity twists the magnetic field lines azimuthally. This generates a magnetic torque in the opposite direction to the rotation, slowing down the rotation and reducing the angular momentum, through the so-called magnetic braking \citep{Allen(b)_2003ApJ...599..363A}. 

Inside $\sim$4$^{\prime\prime}$, the rotation velocity decreases significantly, resulting in a dip in rotation velocity profile at $\sim$1$\overset{\prime\prime}{.}$5 (600 au), roughly consistent with that found in \citetalias{Lee_2016ApJ}. A similar trend is predicted by \citet{Li_2011ApJ}, which explained that in this region, ambipolar diffusion (AD) prevents the magnetic field lines from being dragged inward further (see Section \ref{sec:ambipolar} for details), causing them to accumulate in the post-AD-shock region. The accumulation of magnetic field lines increases the local field-line density and magnetic tension, which not only strengthens magnetic braking and suppresses rotation, but also reduces the radial infall velocity. 

While the efficiency of angular momentum removal via magnetic braking depends on uncertain physical parameters such as magnetic field strength, ionization rate, and dust grain properties \citep{Wurster_2018FrASS...5...39W}, the observed $\sim$95\% reduction in rotation velocity and the accompanying decrease in infall velocity compared to the ballistic model provide empirical constraints on the conditions under which magnetic braking operates effectively.

\subsubsection{Region III: Ambipolar Diffusion}\label{sec:ambipolar}
In Region III ($\sim$1$\overset{\prime\prime}{.}$5 to 0$\overset{\prime\prime}{.}$4, purple shading), the increase in rotation velocity can be attributed to the effects of ambipolar diffusion, which reduces the efficiency of magnetic braking, preventing significant further loss of angular momentum. As the collapse proceeds inward, density increases, leading to a lower ionization fraction and a transition to a region dominated by ambipolar diffusion. In this regime, the coupling between neutral gas and the magnetic field weakens, restricting the inward transport of magnetic flux \citep{Zhao_2016MNRAS.460.2050Z}. Consequently, the magnetic braking effect becomes less efficient, allowing more angular momentum to be retained. 

In this region, the decrease in radial infall velocity is also observed in some simulations \citep[e.g., lower panels of Fig.~5 in][]{Zhao_2020MNRAS.492.3375Z}. As the rotation velocity increases inward, the enhanced centrifugal support counteracts gravity, reducing the infall speed. In addition, magnetic flux is largely removed, rendering the magnetic field dynamically subdominant. The increasing gas density and accumulation near the disk edge lead to further slowdown of infall motion, partly due to momentum conservation. The pressure gradient force also becomes non-negligible in this high-density environment, particularly when the infall motion becomes transonic or subsonic. These effects suggest a deviation from the assumptions of the ballistic model, and Eq.~\ref{eq.5} may no longer be applicable.

The rotation velocity becomes higher than the infall velocity at $\sim$1$^{\prime\prime}$ and thus the envelope starts the final phase of transitioning to the disk and forms the disk at 0$\overset{\prime\prime}{.}$4 (160 au). 

In agreement with \citetalias{Lee_2016ApJ}, the final angular momentum decreases to about one-third of its initial value from $\sim$5$^{\prime\prime}$ to 0$\overset{\prime\prime}{.}$4.

\subsection{Magnetic Field Line Measurement} 

\citet{Yen_2021ApJ...907...33Y} observed polarization in the envelope of this system using the JCMT and found that the mean orientation of the magnetic field at a resolution of $\sim$$14^{\prime\prime}$ is roughly aligned with the rotation axis, with a misalignment of approximately 30$^\circ$. 

The alignment between the magnetic field orientation and the rotation axis significantly influences the efficiency of magnetic braking. Simulations by \citet{Joos_2012A&A...543A.128J} and \citet{Tsukamoto_2015ApJ...810L..26T} demonstrated that when the magnetic field is aligned or only moderately misaligned ($\sim$30$^\circ$–40$^\circ$), magnetic braking can effectively remove angular momentum from the system.

In our case, the specific angular momentum is reduced to roughly one-third of its initial conserved value and is thus consistent with these theoretical expectations. Observationally, \citet{Galametz_2020A&A...644A..47G} found a correlation between the misalignment of magnetic fields with outflow axes and the angular momentum content in low-mass protostars. Their results show that aligned systems tend to exhibit lower angular momentum in the inner envelope, further supporting the idea that magnetic braking is more efficient in aligned configurations.

\section{Conclusion}\label{sec:conclusion}
Using high-resolution ALMA observations, we investigated C$^{18}$O kinematics in the envelope-disk transition region of HH 111. The observed absorption features in the southern and northern regions are consistent with infalling gas, as their velocities are redshifted by $\sim$0.6-0.7 km s$^{-1}$ along the line of sight, which matches the average best-fit infall velocity of $\sim$0.7 km s$^{-1}$ in Region I.

Our best-fit model shows that the infall velocity decreases starting at $\sim$13$^{\prime\prime}$ (5200 au), likely due to magnetic tension by pinched field lines. The rotation velocity increases in the outer envelope, consistent with the angular momentum conservation, but begins to deviate from the conserved value within $\sim$5$^{\prime\prime}$ (2000 au), reaching a minimum at $\sim$1$\overset{\prime\prime}{.}$5 (600 au), as a result of a significant angular momentum loss caused by magnetic braking. Inside 1$\overset{\prime\prime}{.}$5 (600 au), the rotation velocity rapidly increases, likely due to weaker magnetic braking because of ambipolar diffusion of magnetic field lines, eventually reaching Keplerian rotation at 0$\overset{\prime\prime}{.}$4 (160 au). 

Overall, the specific angular momentum decreases to one third of the initial value in the transition region, consistent with previous findings. This study provides a clearer understanding of the velocity profiles in the transition region and highlights the role of magnetic braking in the transport of angular momentum. 

Further constraints on magnetic field strength, orientation, and ionization fraction in this region are essential to quantify the relative contributions of magnetic braking and other non-ideal MHD effects.

%

\vspace{5mm}
\section*{Acknowledgements}
We sincerely thank the editor and reviewer for their kind and constructive feedback. This paper makes use of the following ALMA data: ADS/JAO. ALMA\# 2012.1.00013.S and 2016.1.00389.S. ALMA is a partnership of ESO (representing its member states), NSF (USA) and NINS (Japan), together with NRC (Canada), NSTC and ASIAA (Taiwan), and KASI (Republic of Korea), in cooperation with the Republic of Chile. The Joint ALMA Observatory is operated by ESO, AUI/NRAO and NAOJ. J.-H. L., C.-F. L., and J. A. L.-V. acknowledge support from the National Science and Technology Council (NSTC) of Taiwan through grants 112-2112-M-001-039-MY3, 113-2124-M-001-008, and 114-2124-M-001-015; the Academia Sinica Investigator Award (AS-IA-108-M01); and the Taiwan Astronomical Research Alliance (TARA, formerly TAOvA), supported by NSTC grant 113-2740-M-008-005. TARA is committed to advancing astronomy in Taiwan and paving the way for the establishment of a national observatory. J.-H. L. also acknowledges support from the New Generation Talent Cultivation Scholarship provided by the Department of Physics, National Taiwan Normal University. Z.-Y. L. is supported in part by NASA grant 80NSSC20K0533, NSF grant AST-2307199, and the Virginia Institute of Theoretical Astronomy (VITA). J.-H. L. further thanks Mr. Ayumu Shoshi for valuable discussions and assistance with CASA data processing.

\facilities{ALMA}


\software{Astropy \citep{Astropy_5725236},  
          CASA \citep{CASA_2022PASP..134k4501C}, 
          Matplotlib \citep{Matplotlib_2007CSE.....9...90H}, CARTA \citep{Comrie_2021ascl.soft03031C}, Scipy \citep{Scipy_2020NatAs...4.1158P}
          RADMC-3D\citep{Dullemond_2012ascl.soft02015D}
          }



\clearpage

\bibliography{sample631}{}
\bibliographystyle{aasjournal.bst}


\end{document}